\documentclass[12pt,notoc]{JHEP3}

\usepackage{amsmath,amssymb,euscript,array,cite}

\input{epsf}
\usepackage{epsfig}

\newcommand {\slsh} [1] {\not{\hbox{\kern-2pt${#1}$}}}

\newcommand {\beq} {\begin{equation}}
\newcommand {\eeq} {\end{equation}}
\newcommand {\ber}{\begin{eqnarray*}}
\newcommand {\eer} {\end{eqnarray*}}
\newcommand {\bea}{\begin{eqnarray}}
\newcommand {\eea} {\end{eqnarray}}
\newcommand{\Nfour} {${\cal N}=4\ $}

\newcommand{\None}{${\cal N}=1\ $}

\newcommand{\Dslash}{\,{\raise.15ex\hbox{/}\mkern-12mu D}}
\def\S1{\boldsymbol{R^3 \times S^1}}
\def\C3{\boldsymbol{R^3 \times S^1 \times C^3}}
\def\comp{\boldsymbol{S^3 \times S^1 }}
\def\s3{\boldsymbol{S^3 }}
\def\R4{\boldsymbol{R^4}}
\def\Z2{\boldsymbol{C^3/ (Z_2 \times Z_2)}}
\def\P5{\boldsymbol{AdS_5 \times RP^5 }}
\def\AdS5orbi{\boldsymbol{AdS_5 \times S^5/ \Gamma }}
\def\orbi{\boldsymbol{R^4 \times (C^3 / Z_k) }}

\title{Non-Perturbative Planar Equivalence and the Absence of Closed String Tachyons}
\author{Adi Armoni \\
Department of Physics,\\ Swansea University,\\
Swansea, SA2 8PP, UK.\\
E-mail: {\tt a.armoni@swan.ac.uk}}

\abstract{We consider 'orbifold' and 'orientifold' field theories from 
the dual closed string theory side. We argue that a necessary condition for planar equivalence to hold is the absence of a closed string tachyonic mode in the dual non-supersymmetric string.
We analyze several gauge theories on $\S1$. In the specific case of $U(N)$ theories with symmetric/anti-symmetric fermions ('orientifold field theories') the relevant closed string theory is tachyon-free at large compactification radius (due to winding modes), but it develops a tachyonic mode below a critical radius. Our finding is with agreement with field theory expectations of a phase transition from a C-parity violating phase to a C-parity preserving phase as the compactification radius increases. In the case of $U(N)\times U(N)$ theories with bi-fundamental matter ('orbifold field theories') a tachyon is always present in the string spectrum, at any compactification radius. We conclude that on $\R4$  planar equivalence holds for 'orientfiold field theories', but fails for 'orbifold field theories' daughters of \Nfour SYM and suggest the same for daughters of \None SYM. We also discuss examples of $SO/Sp$ gauge theories with symmetric/anti-symmetric fermions. In this case planar equivalence holds at any compactification radius - in agreement with the absence of tachyons in the string dual.
}
\begin{document}

\section{Introduction}

 Planar equivalence between supersymmetric and non-supersymmetric gauge
theories \cite{Armoni:2003gp,Armoni:2004ub} proved itself as a useful tool in the study of non-perturbative QCD (see \cite{Armoni:2004uu,Armoni:2007vb} for reviews). Besides its important application for QCD, it is an interesting statement in field-theory per-se: a susy/non-susy pair of gauge field theories becomes equivalent in the limit of infinite number of colors.

The first example of a susy/non-susy pair of planar equivalent gauge theories
was suggested by Kachru and Silverstein \cite{Kachru:1998ys}. They claimed that field theories dual to type IIB string theory on $\AdS5orbi$ are conformal and 'equivalent' to (namely have some common sector with) \Nfour super Yang-Mills. This equivalence does not hold (evidence for that can be found in perturbation theory \cite{Adams:2001jb,Armoni:2003va,Dymarsky:2005nc}), although the idea is very inspiring \cite{Strassler:2001fs}. From the string theory side, the reason for the failure of the equivalence is the presence of twisted sector closed string tachyons in the spectrum: the perturbative string vacuum is not the true vacuum of the theory and thus the theory cannot be equivalent to the supersymmetric \Nfour gauge theory. The equivalence suggested by Kachru and Silverstein could have been true, if the spectrum of the orbifolded theory did not contain tachyons. However, it was shown \cite{Armoni:2003va} that on an orbifold of the form $\orbi$, a twisted sector tachyon is always generated when the orbifold breaks supersymmetry. It is possible to project out the closed string tachyon by special orientifolds. Inspired by \cite{Sagnotti:1995ga,Sagnotti:1996qj}, a tachyon free model and planar equivalence between \Nfour and a non-supersymmetric 'orientifold' gauge theory was first suggested in \cite{Angelantonj:1999qg}.

In this paper we argue that a necessary condition for planar equivalence to hold is the absence of tachyons in the dual string side. This condition is also sufficient if the supersymmetric and the non-supersymmetric model admit the same gravity solution. To this end we will analyze several gauge theories on $\S1$ and we will argue that our conditions, formulated in the string side, are {\em equivalent} to the necessary and sufficient conditions that were formulated in pure field theory language \cite{Kovtun:2003hr,Kovtun:2004bz}. Moreover, besides being formal, our analysis will suggest when exactly the equivalence should hold. We will conclude that an orientifold that removes the closed string tachyon from the spectrum is needed. Thus non-supersymmetric theories that live on an orbifold singularity (without any orientifold) cannot be planar equivalent to a supersymmetric theory. In order to project out the tachyon from the spectrum an orientifold is needed. In some cases, namely when the theory is formulated on $\S1$, {\em with a small radius}, even when an orientifold plane is present the equivalence fails --- in agreement with the field theory analysis \cite{Barbon:2005zj,Unsal:2006pj}. The reason is that on $\S1$ a tachyonic mode may re-appear below a certain compactification radius. We will also analyze non-supersymmetric $SO/Sp$ gauge theories on $\S1$. In those cases the string background is supersymmetric and thus free of tachyons. We will see that in this case planar equivalence holds at any compactification radius --- in agreement with the field theory prediction.  

The idea of this paper is not entirely new. A relation between closed string tachyons and noncommutative field theory tachyons for orbifold field theories was suggested in \cite{Armoni:2003va}. In this context the field theory instability is mapped to a string theory instability. A similar relation was obtained recently in \cite{Dymarsky:2005nc}. It is also clear and well known that when the dual string contains a tachyon planar equivalence cannot hold, even for the untwisted sector correlation functions. We suggest here that the absence of tachyons is also a sufficient condition in certain well defined setups and, moreover, it is equivalent to the known field theory conditions. Some of the examples which we discuss in this manuscript are also new.

The organization of the paper is as follows: in section 2 we discuss the conditions for planar equivalence. In section 3,4 and 5 we discuss some examples of planar pairs. Section 6 is devoted to a discussion. In the appendix we show how to the obtain the vacuum structure of the orientifold field theory from open strings.

\section{Necessary and Sufficient Conditions for Planar Equivalence}  

A general theorem for necessary and sufficient conditions of planar equivalence
between a supersymmetric and a non-supersymmetric gauge theories was formulated in \cite{Kovtun:2003hr,Kovtun:2004bz}. Consider a certain 'parent' theory and a projection by a symmetry operation $\Gamma$ ('the orbifold group') that yields a certain 'daughter' theory. The 'daughter' theory is planar equivalent to the 'parent', if the global symmetry $\Gamma $ that was used in the projection is not spontaneously broken. 

Let us suggest a necessary and sufficient condition for planar equivalence in terms of the dual string theory. Since the equivalence is between large-$N$ theories, it translates to an equality of {\em free} (namely $g_{\rm st}=0$) string theories. We are thus after the equivalence of the {\em common} string spectra and the vacuum state of the two theories. The equivalence of the strings spectra is guaranteed by construction. The only dynamical issue is the equivalence of the vacuum state. The only way to invalidate the equivalence is if the perturbative vacuum is not the true vacuum of the theory. This can happen if the non-supersymmetric theory contains tachyons.

Let us see how the field theory conditions, formulated in \cite{Kovtun:2004bz} are mapped to the string side.

Consider a global symmetry which is spontaneously broken. It means that a certain
order parameter ${\cal O}$ obtains an expectation value
\beq
\langle {\cal O} \rangle \ne 0 \, .
\eeq
Consider the case when the operator ${\cal O}$ couples to a tachyon. Planar equivalence cannot hold since the true vacuum of the theory is not the perturbative  vacuum. Thus, it is necessary that there is no tachyon in the spectrum. We can also assert that a tachyon will induce a SSB of ${\cal O}$ in the gauge theory side. 

In order to demonstrate our idea, let us consider a well known example:
the failure \cite{Dymarsky:2005nc} of the equivalence between $U(N)$ \Nfour Super Yang-Mills and the
field theory that lives on D3 branes of type 0 string theory. The latter is a $U(N)\times U(N)$ gauge theory with six scalars in the adjoint of each of the gauge factors and four Dirac bi-fundamental fermions \cite{Klebanov:1999ch}. It can be thought of as a $Z_2$ orbifold projection of \Nfour SYM. This $Z_2$ global symmetry, that acts as the interchanges of the two gauge factors is spontaneously broken in the vacuum.
The order parameter is \cite{Klebanov:1999um}
\beq
\langle {\cal O} \rangle = \langle {\rm tr}\, F_1 ^2 - {\rm tr} \, F_2 ^2 \rangle \ne 0
\eeq
The above operator ${\cal O}$ couples to the type 0 tachyon. Under the $Z_2$ operation $T \rightarrow -T$. We argue that in this case {\em the phenomenon of SSB in the field theory side is identified with a SSB in the string side, as the tachyon field develops a vev.} This statement can be considered as part of the AdS/CFT correspondence dictionary, since field theory operators are identified at large-$N$ with classical fields of the dual string. The full potential of the tachyon field should contain a minimum \cite{Armoni:2005wt} and it should look, qualitatively, as a Higgs potential: $V(T) = M^2 T^2 + \lambda T^4$. The symmetric vacuum $T=0$ is not the true vacuum of the theory. In the true vacuum $T= \pm \sqrt { -{M^2 \over 2\lambda}}$ and the $Z_2$ symmetry $T\rightarrow -T$ is spontaneously broken.

The general picture is as follows: the operator ${\cal O}$ is charged under the 'orbifold' action $\Gamma$ (otherwise it could not be an order parameter). There exists a classical field $T$ in the dual string which couples to ${\cal O}$ and hence this field is also charged under the 'orbifold' $\Gamma$ (this is why it belongs to the twisted sector of the string)
\beq
S = S_{\rm Bulk} + S_{\rm Brane} + S_{\rm Interaction} \, ,
\eeq
where
\beq
 S_{\rm Interaction} = \int d^4 x \,  T{\cal O} + ... \, \label{order} .
\eeq

 When $\langle {\cal O} \rangle \ne 0$ it implies that $\langle T \rangle \ne 0$, namely that $T$ is a tachyon and vice versa: the existence of a tachyon in the dual string implies SSB in the field theory side. Note that ${\cal O}$ need not be necessarily a local operator. For our argument, it is sufficient that a bulk field couples to it. 

 There is another subtle issue that needed to be discussed. It is well known that certain phase transitions are characterized not by the appearance of a tachyon, but by a sudden change in the dual geometry. An example is the confinement/de-confinement phase transition which is described in the bulk by a transition from the thermal AdS solution to a black-hole solution. It is possible that planar equivalence will fail due to a transition in the geometrical description of the non-supersymmetric theory ? The answer is positive\footnote{I am indebted to Asad Naqvi for numerous discussions on this issue.}. Although the gravity equations of both the supersymmetric and the non-supersymmetric theory are identical and one naively expects the same solution for both theories it is possible that the two theories will admit different solutions in cases where the orbifold projection removes the closed string {\em fermions} from the spectrum. The reason is that the presence of fermions obstructs certain solutions and thus their presence in non-supersymmetric theories allows more solutions. We will discuss such an example in section 3. In those cases the absence of tachyons in the spectrum is only necessary, in all other cases it is also sufficient.  
 
In the following sections we will analyze several cases and we will see that our criteria for planar equivalence hold in all examples. Moreover, it will turn out to be a useful tool in deciding which non-supersymmetric gauge theory is a good candidate to be equivalent to a supersymmetric theory, since it is easier to analyze the spectrum of a classical string theory than a strongly coupled gauge theory.
  
\section{Orientifold Field Theories}

Consider type 0B string theory. Let us add $N$ coincident D3-branes and the Sagnotti
O'3 plane. The theory on the branes is the 'orientifold'
daughter of $U(N)$ \Nfour super Yang-Mills. It contains six real scalar in the adjoint representation and four Dirac fermions in the antisymmetric representation. 

Now let us consider the space $\C3$. The worldvolume 
of the branes is $\S1$. If the space was $\R4$ the closed string tachyon of the type 0B string would decouple from the branes, due to the orientifold. However, when the theory is compactified
 on a circle, the closed string tachyon with a non-zero winding number $w$ couples to the brane and its mass becomes 
\beq
\alpha' M^2 = -2 + {(2\pi R w )^2 \over \alpha'}
\eeq

And therefore, when $R^2 < {\alpha ' \over 2\pi ^2}$ the theory becomes tachyonic, due to the mode with $w=1$ and the perturbative string vacuum becomes unstable. Since \Nfour SYM and its non-supersymmetric daughter are conformal the above analysis has no interesting implications for the dual gauge theory. It implies that at zero compactification radius planar equivalence does not hold. It is possible that planar equivalence does not hold at any radius $R < \infty$, if a new gravity solution emerges for the type 0 string. Such a scenario was suggested in \cite{Hollowood:2006cq}.

A similar analysis can be carried out when a stack of $N$ (fractional) D3 branes and an O'3 plane are placed on a $\Z2$ orbifold singularity. The gauge theory on the stack is a $U(N)$ gauge theory with a fermion in the antisymmetric representation \cite{DiVecchia:2004ev}. We can use the above results and express them in field theory quantities (namely replace ${1\over \alpha '}$ with $\Lambda ^2 _{\rm QCD}$) to argue that a tachyonic mode will develop below a critical compactification radius $R_c \sim 1 / \Lambda _{\rm QCD} $. 

The above results are in full agreement with the expectation from the field theory side. Below a certain critical radius planar equivalence for 'orientifold field theories' does not hold \cite{Barbon:2005zj} due to SSB of charge conjugation symmetry \cite{Unsal:2006pj} (see appendix for a stringy derivation of this result). There is an accumulating evidence, both from lattice simulations \cite{DeGrand:2006qb,Lucini:2007as} and from analytic considerations \cite{Hollowood:2006cq,Armoni:2007rf} that C-parity is restored and planar equivalence holds when the compactification radius is sufficiently large.

We suggest that the tachyon vev serves as an order parameter for C-parity breaking in the dual string. Below   $R_c $, $\langle T \rangle \ne 0$ but above $R_c $, the mass of $T$ becomes positive and  $\langle T \rangle = 0$.

\begin{figure}[ht]
\centerline{\includegraphics[width=7cm]{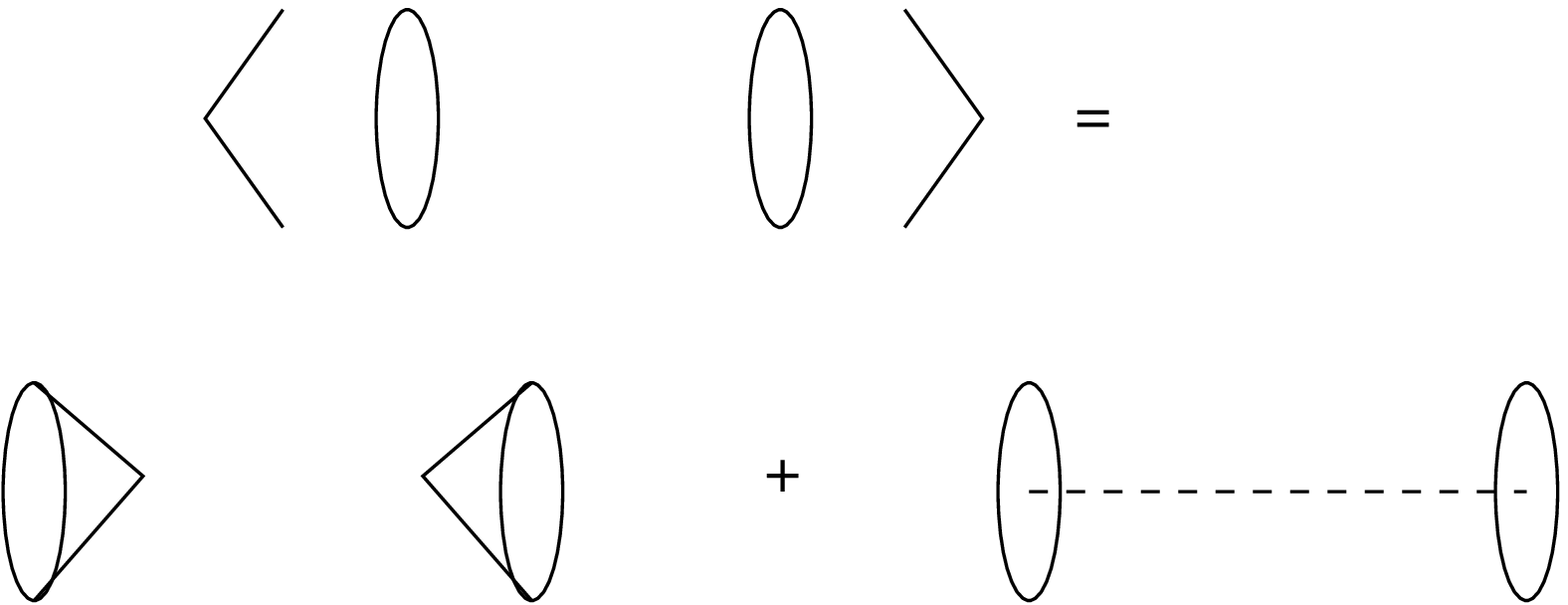}}
\caption{The correlation functions of two well separated Wilson loops is given by the disconnected piece plus a contribution due to an exchange of the lowest glueball in the spectrum.} \label{polyakov}
\end{figure}

In the gauge theory the order parameter for C-parity breaking is the Polyakov loop \cite{Unsal:2006pj}. Since the winding tachyon has the same quantum numbers as the Polyakov loop, it will couple to its worldsheet and thus the general discussion \eqref{order} applies to this case as well, although the order parameter for C breaking is non-local. Another way of seeing that the winding tachyon is the correct object to consider is to note that in QCD the correlation function of two well-separated Wilson loops ($|x| \rightarrow \infty$) is given by the absolute value of a single Wilson loop and a correction due to an exchange of the lowest
glueball in the spectrum
\beq
  \langle W^\dagger (0) W(x) \rangle = |\langle W \rangle|^2 + C \exp -\mu |x| \, .
\eeq
A similar relation $\langle P^\dagger (0) P(x) \rangle = |\langle P \rangle|^2 + C \exp -\sigma |x|$ holds for the Polyakov loop, except that the glueball is replaced by the QCD string. When the Polyakov loop expectation value is zero (the C-parity conserving phase), the correlation function of two well separated Polyakov (or Wilson) loops is controlled by the lowest excitation of the QCD string, whose mass is related (proportional) to the lowest excitation in the tower (see also figure \eqref{polyakov}.).

 In the string description this excitation is the winding tachyon. As we decrease the compactification radius the mass of the 'tachyon' decreases and eventually it becomes zero at $R=R_c$. Below the critical radius we obtain a tachyonic mass and the result will be a flow to a new vacuum where C-parity is broken. Thus our scenario advocates C-parity breaking as a second order phase transition. It predicts a massless 'glueball' and a vanishing string tension at the transition. It will be fascinating if a lattice simulation will confirm this scenario\footnote{It is possible, however, that the phase transition is triggered by a new gravity solution. If this is the case, the phase transition will be of a first order.}. 

\section{Orbifold Field Theories}

Consider type 0B on $\C3$ with D3 branes on $\S1$. As we discussed in section 2, the theory on the brane is the 'orbifold field theory': a $U(N)\times U(N)$ gauge theory with six scalars in the adjoint of each gauge factor and four Dirac bi-fundamental fermions. The bulk contains a closed string tachyon. It couples to the brane at any compactification radius and hence the prediction is that planar equivalence between this theory and \Nfour super Yang-Mills does not hold. This is indeed the case \cite{Dymarsky:2005nc}.

We can consider the orbifold daughter of \None super Yang-Mills theory as well.
It is a $U(N) \times U(N)$ gauge theory with one Dirac fermion in the bi-fundamental representation. The realization of this theory is very similar to the previous theory: it 'lives' on the D3 branes of type 0B string theory, placed on $\Z2$ orbifold singularity. In this case as well there is a closed string tachyon in the bulk, at any compactification radius, and thus the prediction is that planar equivalence between this theory and \None super Yang-Mills does not hold. For small compactification radius this is indeed the case \cite{Tong:2002vp}. On $\R4$ it is more difficult to confirm the failure of planar equivalence, but there are several hints. The domain walls of the daughter theory are non-BPS objects \cite{Armoni:2005wt} although the domain walls  of \None are\footnote{A different point of view was advocated in \cite{Kovtun:2005kh}.}. Another hint comes from the noncommutative version of the theory: for any value of the noncommutativity parameter $\theta$ the perturbative vacuum of the theory becomes unstable and in fact the instability in the gauge theory can be related to the bulk tachyons \cite{Armoni:2001uw}. If we could connect the noncommutative theory to the commutative theory, namely to take the limit $\theta \rightarrow 0$, it would be sufficient to disprove planar equivalence. Unfortunately, it is known that the limit $\theta \rightarrow 0$ is singular and therefore it is not clear whether the noncommutativity deformation can teach us about the commutative theory.\footnote{There are also contrary hints: recently \cite{Unsal:2007fb} this gauge theory was analyzed on $\comp$ (with a small $\s3$ radius) and a transition from a $Z_2$ broken phase to a $Z_2$ preserving phase was found and thus it was suggested that planar equivalence may hold on $\R4$. Note however that the analysis was carried out for the free theory whereas on $\R4$ a $Z_2$ restoration should be a non-perturbative effect. Therefore, it is not clear whether we can draw a lesson about the theory on $\R4$ from this calculation.}

 To conclude this section: we argue that 'orbifold gauge theories' cannot be planar equivalent to supersymmetric theories, since their string theory description involves a closed string tachyon. Unfortunately, in the case of \None SYM and its orbifold daughter we cannot make a firm prediction: we use a flat space analysis and ignore possible effects due to R-R flux. we cannot exclude a scenario where the string dual of our gauge theory contains a R-R flux whose coupling to the tachyon will shift its mass towards a positive value \cite{Klebanov:1998yy}.

\section{$SO(N)$ and $Sp(N/2)$ theories}

There is another class of interesting non-supersymmetric gauge theories that becomes planar equivalent to supersymmetric theories.

Consider D3 branes in type IIB string theory. Let us place an $O3$ plane on top of the D3 branes. The theories on the branes are supersymmetric \Nfour $SO/Sp$ gauge theories. Both theories are dual to type IIB string theory on $\P5$ \cite{Witten:1998xy}. Both contain the same spectrum of unoriented closed strings at large-$N$ \cite{Witten:1998xy} and moreover are also related to the \Nfour $SU(N)$ theory by a $Z_2$ truncation.

There is, however, another interesting option. We can place anti D3-branes instead of D3-branes. The gauge theories on the branes are either a non-supersymmetric $SO(N)$ theory with symmetric fermions, or a $Sp(N/2)$ theory with antisymmetric fermions. These models were constructed first by Sugimoto \cite{Sugimoto:1999tx}. Since the orientifold is a $1/N$ effect, it was suggested that at infinite $N$ the non-supersymmetric theory become actually supersymmetric \cite{Uranga:1999ib}. This is a simple version of planar equivalence. In terms of representation theory the statement is that the symmetric and the antisymmetric representations become indistinguishable at infinite $N$. Indeed, all the Casimirs of these representations coincide at infinite $N$. 

This form of planar equivalence is quite weak and does not involve any non-perturbative dynamics. It is a simple statement about representations and therefore it holds at any compactification radius. This is in perfect agreement with our criterion of the absence of tachyons in the bulk theory. Since the bulk is type IIB, namely a supersymmetric string, there are no closed string tachyons. So, we could predict the validity of planar equivalence from the bulk dynamics. Unfortunately, in this case it is a trivial prediction.

\section{Summary}

In this paper we formulated necessary and sufficient conditions for planar equivalence between a large-$N$ supersymmetric and non-supersymmetric theories. Our criteria are the absence of closed string tachyons in the dual non-supersymmetric string spectrum and the requirement that the supersymmetric and the non-supersymmetric theories will admit the same classical gravity solution.

These criteria successfully reproduce known results for the daughters of \Nfour and \None super Yang-Mills. In some cases it becomes a highly non-trivial prediction for strongly coupled non-supersymmetric gauge theories: in particular we predict the success of planar equivalence between the 'orientifold' daughter of \None SYM and \None SYM itself on $\R4$ and a second order phase transition at a critical radius on $\S1$. We also argue that the 'orbifold' daughter of \None SYM is not planar equivalent to \None SYM.

It will be interesting to find other situations where spontaneous symmetry breaking is accompanied by the presence of tachyons. It was recently argued that chiral symmetry breaking in the probe limit is dual to {\em open} string tachyon condensation \cite{Casero:2007ae}. We suggest that chiral symmetry breaking in the Veneziano limit ($N_c \rightarrow \infty$, $Nf/Nc={\rm fixed.}$) might be described by closed string tachyon condensation.

{\bf Acknowledgements:} I would like to thank C. Angelantonj, E. Imeroni, T. Hollowood, B. Lucini, A. Naqvi, M. Shifman, M. Unsal, A. Uranga and G. Veneziano for useful discussions and comments. I am supported by the PPARC advanced fellowship award.

\section{Appendix: The vacuum structure of the 'orientifold field theory' on $\S1$ from winding open strings}

We are interested in the vacuum structure of the compactified 'orientifold field theory' (a $U(N)$ gauge theory with an antisymmetric Dirac fermion). Our starting point is the Hanany-Witten brane configuration of ref.\cite{Armoni:2003gp}. We have an $O'4$ plane and $N$ D4-branes suspended between
 two NS fivebranes. Now compactify the $X^3$ direction on circle of radius $R$.
The D-branes and the orientifold plane wrap the $X^3$ direction. Next perform a T-duality along the $X^3$ direction\footnote{T-duality in type 0 string theory was argued in ref.\cite{Bergman:1999km}}. We obtain a circle of radius $R' ={1 \over R}$ (we set $\alpha' =1$). There are two orientifold planes at the fixed points $V=0$ and $V=\pi$. In addition there are localized D3-branes at arbitrary angles $V_1,V_2,...,V_N$ and their mirrors at $-V_1,-V_2,...,-V_N$. Note that the D-branes are localized at the interval $[0,\pi)$.

 \begin{figure}[ht]
\centerline{\includegraphics[width=8cm]{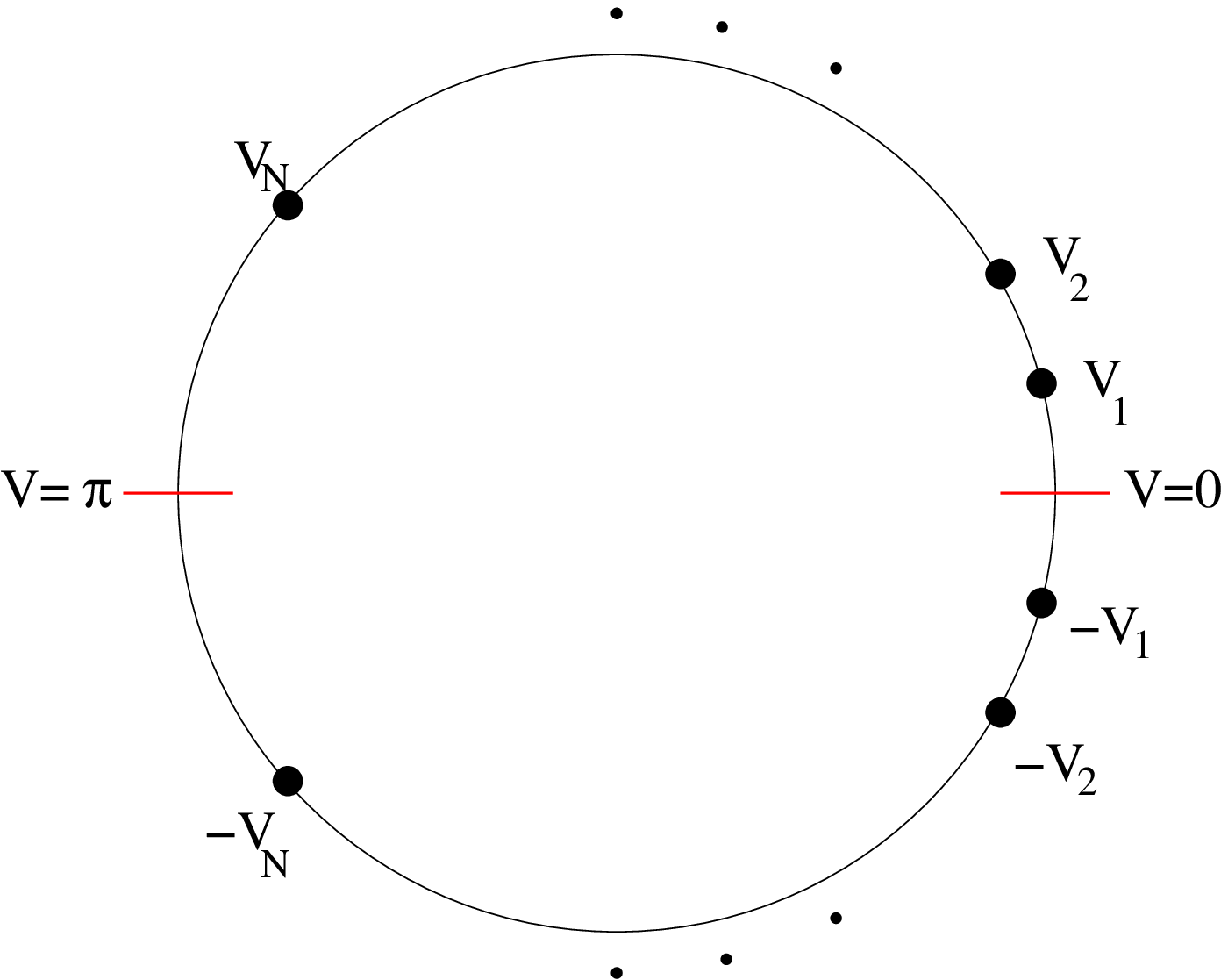}}
\caption{The brane configuration after T-duality. There are orientifold planes at $V=0$ and $V=\pi$. The D-branes are localized at $V_1,V_2,...,V_N$ and at the mirror $-V_1,-V_2,...,-V_N$.} \label{circle}
\end{figure}

The mass of an open string that does not cross the orientifold (a boson) is $(V_i-V_j) R'$. The string can wind around the circle. If the winding number is $n$ the mass is
\beq
m^{(n)} _{\rm boson} = (2\pi n + V_i-V_j)R'
\eeq
$n$ can be either positive or negative. 
Similarly the mass of open strings that cross the orientifold odd number of times (fermions) is
\beq
m^{(n)} _{\rm fermion} = (2\pi n + V_i+V_j)R'
\eeq

Let us calculate the total vacuum energy. It is $E=\sum \omega _B - \sum \omega _F$ of the open strings modes, namely
\bea
& &
E(V_1,V_2,...,V_N) =  \\
& &
 \sum _n \int {d^3 k \over (2\pi)^3} \left ( \sum _{i,j} \log (k^2 + m^2 _{\rm boson} ) - \sum _{i \ne j} \log (k^2 +m^2 _{\rm fermion} ) \right ) \, ,
\nonumber
\eea
or
\bea
& &
E(V_1,V_2,...,V_N) =  \label{energy} \\
& &
 \sum _n \int {d^3 k \over (2\pi)^3} \left ( \sum _{i,j} \log (k^2 + R'^2 (2\pi n + V_i-V_j)^2 ) - \sum _{i \ne j} \log (k^2 + R'^2(2\pi n + V_i+V_j) ^2) \right ) \, . 
\nonumber
\eea

\begin{figure}[ht]
\centerline{\includegraphics[width=8cm]{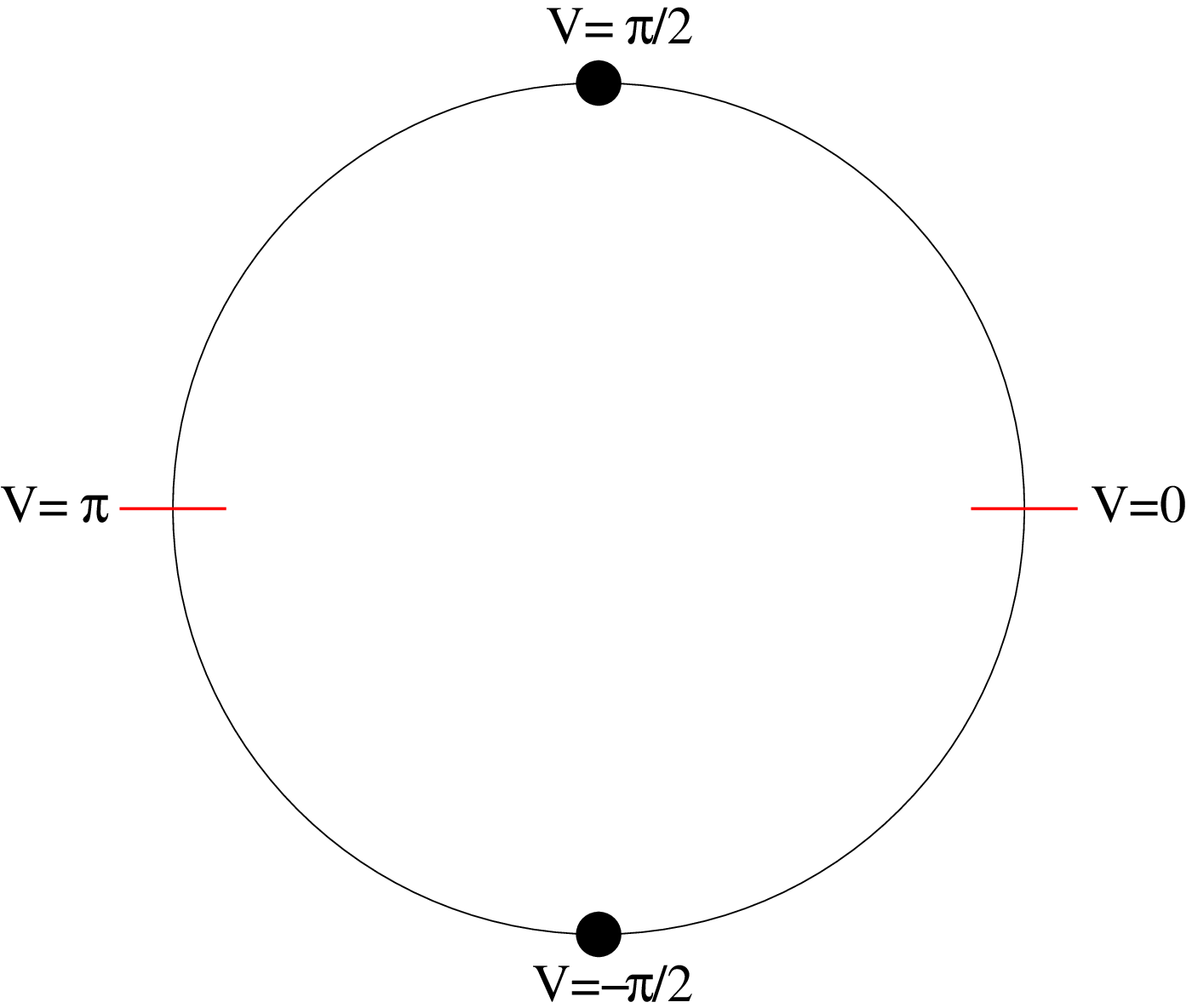}}
\caption{The vacuum configuration. There are $N$ coincident D-branes  at $V=\pi/2$ and their mirrors at $V=-\pi/2$.} \label{vacuum}
\end{figure}

 The above expression \eqref{energy} is identical to the expression that was derived in ref.\cite{Unsal:2006pj}. The minimum of the potential is when all the branes coincide at $V_1=V_2=...=V_N={\pi \over 2}$ and their mirrors at $V_1=V_2=...=V_N=-{\pi \over 2}$, see figure \eqref{vacuum}.

Note that since the positions of the branes are identified with v.e.v.'s the fermions becomes massive through the term 
\beq
{\cal L}=\Phi \bar \Psi \Psi
\eeq
($\Phi$ is the zero mode of $A^3$). The result is a $U(N)$ gauge theory with a massive fermion, $m=\pi R' = {\pi \over R}$. Clearly it cannot be equivalent to \None SYM on a circle, where in the vacuum the branes are distributed uniformly along the circle and the gauge symmetry is broken to $U(1)^N$.

\end{document}